\documentclass[12pt]{article}
\usepackage[utf8]{inputenc}
\usepackage{natbib}
\usepackage{setspace}
\usepackage[left=2cm,right=2cm,top=2cm,bottom=2.5cm,bindingoffset=0cm]{geometry}

\newcommand{\etal}{{et al.}}

\begin{document}

%\noindent ???~524.38, 524.383, 521.1

%\large
{\it
\noindent
Yu.\,T.\,Tsap, A.\,V.\,Stepanov,  Yu.\,G.\,Kopylova\\
``On the Description of Transverse Wave Propagation Along Thin Magnetic Flux Tubes'' \\
Geomagnetism and Aeronomy, 2018, V.~58, N~7, pp.942--946.\\}
DOI: 10.1134/S001679321807023X

\bigskip

\begin{center}
{\large ON THE DESCRIPTION OF TRANSVERSE WAVE PROPAGATION ALONG THIN MAGNETIC FLUX TUBES}

\smallskip
{Yu.\,T.\,Tsap$^{a,b,*}$, A.\,V.\,Stepanov$^{b,**}$,  Yu.\,G.\,Kopylova$^{b,***}$\\}

\medskip
{\it \normalsize
$^{a}$ Crimean Astrophysical  Observatory of the Russian Academy of Sciences, Crimea, Russia\\
$^{b}$ Pulkovo Observatory of the Russian Academy of Sciences, St. Petersburg, Russia\\
$^{*}$email: yur\_crao@mail.ru, $^{**}$email: stepanov@gaoran.ru, $^{***}$email: yul@gaoran.ru
}
\end{center}

%\smallskip
%\doublespacing

\vspace{-10mm}
\begin{abstract}
Two approaches are used for description of linear transverse
(kink) modes excited in a vertical thin magnetic flux tube. First
one is based on the elastic thread model~\citep{Spruit81}. The
second one  follows from the the Taylor and Laurent series
expansions of wave variables with respect to the tube radius
inside and outside of the magnetic flux
tube~\citep{Lopin&Nagorny13}. It has been shown that the main
reason of the discrepancy of these approaches is related to the
phenomenological equation of plasma motion used in the former
case. This suggests that results obtained on the basis of this
equation should be revised.
\end{abstract}

%\keywords{Sun: magnetic fields $\cdot$  Sun: oscillations $\cdot$ Sun: photosphere}

\section{Introduction}
\label{chap1}

It is well known that the magnetic  flux tubes formed under action
of convective motions can play an important role in the solar
atmosphere. According to existing concepts the tubes with
diameters of about 100~km or less are the main structural elements
of the solar photosphere~\citep{Stenflo11,Ji et
al.12,Sharykin&Kosovichev14} and they can contain up to 90 per
cent of the total magnetic field~\citep{Howard&Stenflo72,Ruedi
etal.92}. These magnetic structures transmit the convective energy
of plasma motion from the photosphere to the corona like
waveguides. The last property is the subject of investigation
presented in this article.

 It is well
established now that the high frequency Alfv\'en type modes (kink,
torsional) generated in the photosphere are strongly damped in the
chromosphere \citep{De Pontieu et al.01,Leake etal.05, Soler
etal.17} if their periods $T < 10$~s while the waves with  $T >
40$~s are quite strongly reflected in the transition
region~\citep{Tsap06, Soler etal.17}. This suggests that waves
with $T = 10$--40~s can be responsible for  heating  at least of
lower corona~\citep[see also][]{Gelfreikh etal.04}. In turn, there
are many indications that linear kink  waves with $T > 1$~min
excited in magnetic flux tubes  of the solar
photosphere~\citep{Fujimura&Tsuneta09, Jess et al.17} are a prime
candidate for heating and accelerating the fast wind because they
have the ability to transport energy over large distances in the
corona~\citep{Morton et al.15}. However, the possibility of
effective generation and propagation of these waves in the
photosphere is still unclear.

The theory of linear kink modes based on the thin magnetic flux
tube approximation for the first time was proposed independently
by \citet{Ryutov&Ryutova76} and \citet{Spruit81}. Since the
approach of \citet{Ryutov&Ryutova76}  suggests the constant cross
section of the tube we will not discuss  it further in detail. The
approach proposed by \citet{Spruit81} is more general and it is
often used up to now to description of magnetic flux tube dynamics
in the solar stellar convection zone
\citep{Fan.09,Weber&Browning16}. Note that the thin flux tube
approximation can also describe MHD waves in the solar photosphere
and lower chromosphere~\citep{Lopin&Nagorny17} quite well.

One of the main problem of this approach connected with the
phenomenological equation of motion. Fist of all, this concerns
the effect of the enhanced inertia caused by the backreaction of
the the ambient medium to the relative motion of the flux
tube~\citep{Spruit81}. The correct treatment of this contribution
is a source of continuing controversy \citep{Choudhuri90, Cheng92,
Fan etal.94, Moreno-Insertis etal.96, Longcope&Klapper97, Osin
etal.99}.

In order to describe the MHD waves in the thin magnetic flux tubes
approximation  in addition to the Spruit's approach  the second
method based on the power series expansion of wave variables with
respect to the tube radius \citep[see also][]{Roberts&Webb78,
Ferriz-Mas at al.89} is used. Recently, \citet{Lopin etal.14},
based on the system of linearized ideal MHD equations and using
this method have shown that the phenomenological equation of
motion can be used for the description of kink modes in thin
magnetic flux tubes if we take into account the radial component
of the magnetic flux tube in the  induction equation. This means
that the phenomenological equation of motion  can be used for the
description of MHD waves excited in thin magnetic flux tubes  when
the relative motion of the ambient medium is equal to zero.

This study is devoted to the analysis of the phenomenological
equation of motion proposed by proposed by~\citet{Spruit81} based
on the the power series expansion of wave variables for linear
kink mode propagation in a vertically stratified thin magnetic
flux tube.

\section{Vertical non-twisted magnetic flux tubes in stratified \\ atmosphere and their properties}
\label{chap2}

Let us first slightly generalize results obtained earlier for a
vertical isolated magnetic flux tube in  the stratified atmosphere
with the radius $a$~\citep{Roberts&Webb78,Hollweg81, Ferriz-Mas at
al.89}.

The equilibrium magnetic field of  an axisymmetric untwisted flux tube in cylindrical coordinates can be written as
\begin{equation}
\label{tube field} {\bf B} = B_z(z) {\bf e}_z + B_r(r,z) {\bf e}_r,
\end{equation}
where  ${\bf e}_z$ and ${\bf e}_r$ are the unit vectors.

For the sake of simplicity we can assume that the radial component of its magnetic field due to the tube expansion with height is
\begin{equation}
\label{alpha}
 B_r(r,z)= \alpha r,\;\;\; \alpha = \mathrm{const}.
\end{equation}
This suggestion can be considered as some generalization of the
thin flux tube approximation since $\alpha$ can take arbitrary
values. Note that in the thin magnetic flux tube approximation
$B_z(r,z)$ and $B_r(r,z)$ components of the tube magnetic field
are expanded in a Taylor series~\citep[e.g.,][]{Ferriz-Mas at
al.89, Lopin&Nagorny13}
\begin{equation}
\label{Bz expansion} B_z(r,z)=  B_z(0,z) + r^2\frac{\partial B_r^2}{\partial r^2}(0,z),
\end{equation}
\begin{equation}
\label{Br expansion} B_r(r,z)=  B_r(0,z) + r\frac{\partial B_r}{\partial r}(0,z),
\end{equation}
where $B_z(0,z)\equiv B_z(z)$ is independent of $r$ and
$B_r(0,z)=0$ in the lowest order of the series expansion, and
$|B_r(a,z)|/|B_z(z)|  \ll 1$.

The equation of the magnetic field lines, in view of equation~(\ref{alpha}), can be reduced to the relationships
\begin{equation}
\label{magnetic line equation} \frac{dz}{dr}=\frac{B_z}{B_r} = \frac{B_z}{\alpha r}.
\end{equation}
Since the magnetic field lines are closed curves ($\nabla \cdot {\bf B}=0$),  using equation~(\ref{alpha}), we get
\begin{equation}
\label{diver}
\nabla\cdot{\bf B} = \frac{1}{r}\frac{\partial (rB_r)}{\partial r} + \frac{\partial B_z}{\partial z}  =
2 \alpha + \frac{\partial B_z}{\partial z} = 0.
\end{equation}
Hence,  equations  (\ref{tube field}), (\ref{alpha}), and (\ref{diver})~\citep{Roberts&Webb78,Ferriz-Mas at al.89} give
\begin{equation}
\label{radial field}
B_r(r,z)=- \frac{r}{2}\frac{d B_z(z)}{dz}.
\end{equation}

Combining equations (\ref{magnetic line equation}) and~(\ref{diver}), we obtain the differential equation in the form
\begin{equation}
\label{lines}
2 \frac{dr}{r} + \frac{d B_z}{B_z} = 0.
\end{equation}
The general solution of equation (\ref{lines}) can be represented as
\begin{equation}
\label{flux conservation}
B_z  r^2 = \mathrm{const}.
\end{equation}

Equation (\ref{flux conservation}) means  that  the longitudinal
magnetic flux $F_z = B_za^2$ of the thin or wide tube is
conserved. In particularly, this important circumstance was used
by~\citet{Hollweg81} \citep[see also][]{Hollweg84}
and~\citet{Routh etal.07} for description of torsional modes in a
wide magnetic flux tube. However, in the future we shall confine
ourselves to the study of thin flux tubes.

The plasma density in stratified atmosphere decreases with the
hight $z$ inside and outside ($e$) the tube for the isothermal
atmosphere according to the barometric formulae
 \begin{equation}
\label{density}
\rho = \rho_{0} e^{-z/H},\;\;\;\rho_{e} = \rho_{0e} e^{-z/H},
\end{equation}
where $H$ is a characteristic scale hight. In turn, the balance of total pressures, which follows from the integration of the MHD
equilibrium equation, for an isolated ($B_e = 0$) magnetic flux tube is
\begin{equation}
\label{balance}
 p + \frac{B^2}{8\pi} =  p_e.
\end{equation}
Equations  (\ref{density}) and (\ref{balance}) give
\begin{equation}
\label{field}
B_z \propto e^{-z/2H},\;\;\; B_r \propto e^{-z/2H}.
\end{equation}
Hence, equations  (\ref{radial field}) and (\ref{field}) implies that the radial component  of a magnetic flux tube  is
\begin{equation}
\label{radial field+} B_r(r,z)= \left(\frac{r}{4H}\right)B_z(r,0)
e^{-z/2H}.
\end{equation}

Equations (\ref{flux conservation}), (\ref{field}), and
(\ref{radial field+}) describe the dependence of the magnetic
field on $r$ and $z$ in thin vertical magnetic flux tubes.
Besides, equation~(\ref{radial field+}) suggests that the thin
magnetic flux tube approximation ($B_r \ll B_z$) is fulfilled if
the tube radius ($a \ll 4H$). For example, taking in the
photosphere  $H = 150$~km, we find $a \ll 600$~km \citep[see
also][]{Lopin&Nagorny17}.

\section{The phenomenological equation of motion and \\the elastic thread model proposed by Spruit~(1981)}
\label{chap4}

 The approach proposed by \citet{Spruit81} is based the linearized  phenomenological equation of  motion, which using
 standard notation can be represented as
\begin{equation}
\label{motion-ph}
(\rho + \rho_e) {d{\bf v_\perp}/dt} = \delta {\bf F}_\perp.
\end{equation}
Equation~(\ref{motion-ph}) seems to be the  main disadvantage of Spruit' s approach. Indeed,  the transverse force $\delta{\bf F}_\perp$
on the right-hand side of equation~(\ref{motion-ph}) includes the action of the internal force only while the left-hand side corresponds to
the plasma inertia inside and outside ($e$) the magnetic flux tube.

\citet{Spruit81} took formally into consideration external forces using the the pressure balance equation for disturbed quantities
\begin{equation}
\label{disturbed balance}
\delta P =  \delta p_e,
\end{equation}
where the total pressure $P =  p + B^2/(8\pi)$. Indeed, applying
the nabla differential operator to equation~(\ref{disturbed
balance}), in view of the equilibrium equation outside the tube,
$\nabla \delta p_e = \delta \rho_e{\bf g}$, we get
\begin{equation}
\label{spruit}
\nabla \delta P = \nabla \delta p_e =  \delta \rho_e{\bf g}.
\end{equation}
Equation (\ref{spruit}) describes the coupling between internal
and external forces since the force $\delta {\bf F}_\perp$
 (see equation~(\ref{motion-ph})) includes the term $\nabla_\perp \delta P$ and, hence, the external gravity force $\delta \rho_e{\bf g}_\perp$.

\citet{Musielak&Ulmschneider01} in order to prove the consistency of the phenomenological equation~(\ref{motion-ph}) and the
magnetohydrodynamic principles instead of equation~(\ref{spruit}) used the following boundary conditions
\begin{equation}
\label{musielak}
\nabla \delta P = \nabla \delta p_e = \delta \rho_e{\bf g} - \rho_e \frac{\partial {\bf v}_{\perp e}}{\partial t}.
\end{equation}
Equation (\ref{musielak}) differs from equation~(\ref{spruit}) by the inertia term $\rho_e \partial {\bf v}_{\perp e}/\partial t$.
As a result,~\citet{Musielak&Ulmschneider01} are forced to suppose that the velocities normal to the interface within
(${\bf v}_\perp$) and outside (${\bf v}_{\perp e}$) the tube have different signs, i.e. ${\bf v}_\perp = - {\bf v}_{\perp e}$ that
doesn't make sense (in particular, it means that authors used different systems of coordinates inside and outside the tube).

In order to investigate the linear kink oscillations of the vertical  magnetic flux tubes the elastic thread model was elaborated by~\citet{Spruit81}.
This model based on equation~(\ref{motion-ph}) and the thin magnetic flux tube approximation described in Section~(2) gives the following
wave equation~\citep{Spruit81}
\begin{equation}
\label{dispersion-Spruit}
\frac{\partial^2 s_{\perp}}{\partial z^2} - \frac{1}{2 H} \frac{\partial s_{\perp}}{\partial z} +
\frac{\omega^2}{c_k^2} s_{\perp} = 0,
\end{equation}
where $s_\perp$ is the displacement perpendicular to the tube
axis~$z$. The solution of this differential equation is
\begin{equation}
\label{dispersion-Spruit-solaution} s_\perp \propto e^{1/(4H) z
\pm i \sqrt{\omega^2 - \omega_c^2}/c_k z}.
\end{equation}
It describes the propagation of kink waves  with the cutoff
frequency $\omega_c = c_k/2H$.

\section{Power series expansion and  thin magnetic flux\\ tube approximation}
\label{chap3}

Let us consider briefly  main features of the approach,  which  is
based on the method of the power series expansion with respect to
$r$ \citep[see also][]{Roberts&Webb78, Ferriz-Mas at al.89}.

According to this method, the perturbed quantities $f(r, z, t)$ inside a thin magnetic flux tube, using standard notation, can be
expanded in a Taylor series as
\begin{eqnarray}
\label{f-expansion}
f(r,z,t) =  && f(0,z,t) + r \frac{\partial f}{\partial r} (0,z,t) +  r^2 \frac{\partial f}{\partial r} (0,z,t) +\ldots = \cr\cr &&
f_0(z,t) + f_1(r,z,t) + f_2(r,z,t) +\ldots,
\end{eqnarray}
where $|f_{n+1} (z, t)|/|f_n (z,t)| = \mu \ll 1$ and $f(r,z,t)$ is a slowly varying function of  $z$ and~$\varphi$, i.e.
$$
\left|\frac{\partial f}{\partial r}\right| \sim
\frac{1}{\mu}\left|\frac{\partial f}{\partial z} \right| \sim
\frac{1}{\mu r} \left|\frac{\partial f}{\partial \varphi} \right|.
$$
Note that \citet{Lopin&Nagorny13} used a Laurent series outside the tube since  perturbed quantities outside a magnetic flux tube
decrease with $r$ in this region~\citep[e.g.,][]{Tsap et al.01}.

Perturbed quantities inside a thin magnetic flux tube can be represented as
\begin{equation}
\label{s_r}
s_{r}(r, \varphi, z) =  s_{r0} (\varphi, z) + s_{r2} (r, \varphi, z) + \ldots,
\end{equation}
\begin{equation}
\label{s_phi}
s_{\varphi}(r, \varphi, z) = s_{\varphi 0} (\varphi, z) + s_{\varphi 2} (r, \varphi, z) + \ldots,
\end{equation}
\begin{equation}
\label{s_z}
s_{z} (r, \varphi, z)  =  s_{z1} (r, \varphi, z)  + s_{z3} (r, \varphi, z)+ \ldots,\;\;\;
\end{equation}
\begin{equation}
\label{expansion}
 b_{r} = b_{r0}(\varphi, z) +  b_{r2} (r, \varphi, z)  +
 \ldots,\;\;\; b_{\varphi } = b_{\varphi 0}(\varphi, z) +  b_{\varphi 2} (r, \varphi, z)  + \ldots,
\end{equation}
\begin{equation}
\label{expansion b_z}
b_{z} =  b_{z1} (r, \varphi, z) +  b _{z3} (r, \varphi, z) + \ldots,
\end{equation}
\begin{equation}
\label{expansion  P}
\delta P = \delta P_{1} (r, \varphi, z) + \delta P _{3} (r, \varphi, z) + \ldots,
\end{equation}
where the perturbed total pressure $\delta P = \delta p + ({\bf b B})/4\pi$. In turn, perturbed quantities outside a magnetic flux
tube are described by Hankel or Macdonald functions~\citep[e.g.][]{Tsap et al.01}. This suggests that the equation of motion
outside  the tube has the form \citep{Lopin&Nagorny13,Lopin etal.14}
\begin{equation}
\label{motion outside}
\rho_e\frac{\partial^2 {s}_{r0}}{\partial t^2} = \frac{\delta p_{e0}}{r}.
\end{equation}

Thus, using Equations (\ref{tube field}),  (\ref{radial field}), (\ref{balance}), (\ref{radial field+}), (\ref{s_r})--(\ref{motion outside}),
and a system of linear equations of ideal MHD~\citep[e.g.][]{Lopin&Nagorny13}, we can find the wave equation of kink
modes~\citep{Lopin&Nagorny13,Lopin etal.14}
\begin{equation}
\label{dispersion}
\frac{\partial^2 s_{r0}}{\partial z^2} -
\frac{1}{2 H} \frac{\partial s_{r0}}{\partial z} +
\left(\frac{\omega^2}{c_k^2} + \frac{1}{16H^2}\right)s_{r0} = 0,
\end{equation}
where the phase velocity $c_k= B_z/\sqrt{4\pi (\rho + \rho_e)}$. It is easy  to show  that the solution of equation~(\ref{dispersion})
can be represented as~\citep{Lopin&Nagorny13,Lopin etal.14}
\begin{equation}
\label{dispersion-solaution}
 s_{r0} \propto e^{1/(4H)z \pm i\omega/c_k z}.
\end{equation}
Expression~(\ref{dispersion-solaution}) as distinguished from
(\ref{dispersion-Spruit-solaution})
 corresponds to the cutoff
free propagation of kink modes.

\section{ Lopin et al.~(2014) vs Spruit~(1981)}
\label{chap5}

Using equation~(\ref{motion-ph}), \citet{Spruit81} has shown that equation of motion perpendicular to the tube axis
in curvilinear coordinates has the form
\begin{equation}
\label{spruit motion}
(\rho + \rho_e)\frac{\partial^2 {{\bf s}_\perp}}{\partial t^2} = \frac{B^2 +
B_e^2}{4\pi}{\mathbf{c}} +(\rho - \rho_e)({\bf e}_l\times{\bf g})\times{\bf e}_l,
\end{equation}
where ${\bf e}_l$ is the local unit vector oriented along the magnetic-field lines and $\mathbf{c}= ({\bf e}_l\cdot \nabla){\bf e}_l$
is the curvature vector.

\citet{Lopin etal.14} in order to reveal disadvantages of Spruit's approach reduces equation~(\ref{spruit motion})
to equation~(\ref{dispersion}) obtained by \citet{Lopin&Nagorny13} in terms of the linear induction equation in Cartesian
coordinates~\label{lopin}
\begin{equation}
\label{incorrect induction}
b_{x0} = \left(B_z\frac{\partial s_{x0}}{\partial z} + \frac{\partial B_z}{\partial  z}s_{x0} \right)  + \frac{B_x}{x}s_{x0},
\end{equation}
where $b_{r0} = b_{x0}\cos\varphi$ and  $s_{r0} = s_{x0}\cos\varphi$~\citep{Lopin etal.14}. Note that we replace the
subscript $x$ by $x_0$ as distinguished from the corresponding equation obtained by \citet{Lopin etal.14} in order to stress
that we consider zeroth-order quantities.

In view of equation~(\ref{radial field}), we have \citep[see also][]{Lopin etal.14}
\begin{equation}
\label{B_x}
B_x(x,z) = -\frac{x}{2}\frac{\partial B_z}{\partial  z},
\end{equation}
and  after substitution of equation~(\ref{B_x}) to the expression~(\ref{incorrect induction}) the inclination of the perturbed tube axis is
\begin{equation}
\label{l_x-Lopin}
l_x' = \frac{b_{x0}}{B_z} = \frac{\partial s_{x0}}{\partial z} + \frac{s_{x0}}{2B_z}\frac{\partial B_z}{\partial z}.
\end{equation}
Equation (\ref{l_x-Lopin}) does not coincide with the corresponding expression obtained by~\citet{Spruit81} which can be written as
\begin{equation}
\label{l_x-Spruit}
 l_x' = \frac{b_{x0}}{B_z} = \frac{\partial s_{x0}}{\partial z}.
\end{equation}

\citet{Lopin etal.14} have shown that equation~(\ref{spruit
motion}) is reduced to wave equation~(\ref{dispersion}) after
substitution equation~(\ref{l_x-Lopin}) to equation~(\ref{spruit
motion}). Whence, comparing equations~(\ref{l_x-Lopin})
and~(\ref{l_x-Spruit}), they concluded that the main reason of the
discrepancy  is associated with the radial component of  the  tube
magnetic field $B_r$ which was not taken into account by
\citet{Spruit81}. We do not agree with this inference. Really, it
suggests that the approach proposed by \citet{Lopin&Nagorny13} can
be considered as a generalization of Spruit's
results~\citep{Spruit81} and at
\begin{equation}
\label{Spruit-Lopin inequality}
\left|\frac{1}{s_{x0}}\frac{\partial s_{x0}}{\partial z}\right|
\ll \left|\frac{1}{2B_z}\frac{\partial B_z}{\partial z} \right| = \frac{1}{4H},
\end{equation}
Equation~(\ref{dispersion}) obtained by \citet{Lopin&Nagorny13} \citep[see also][]{Lopin etal.14} should be reduced  to Spruit's
equation~(\ref{dispersion-Spruit}). However, if we take into account that $s_{r0} = s_{x0}\cos\varphi$ and inequality~(\ref{Spruit-Lopin inequality}),
the wave equation~(\ref{dispersion}) obtained by \citet{Lopin&Nagorny13} takes the form
\begin{equation}
\label{dispersion-S+L}
\frac{\partial^2 s_{x0}}{\partial z^2} + \left(\frac{\omega^2}{c_k^2} +\frac{1}{16H^2}\right)s_{x0} = 0,
\end{equation}
where we adopt the displacement $s_{x0} = s_\perp$ since $s_{y0}
\equiv 0$~\citep{Lopin etal.14}. By comparing
equations~(\ref{dispersion-Spruit}) and~(\ref{dispersion-S+L}), we can
conclude that they are not coincided. Thus, the discrepancy
between results obtained by \citet{Spruit81} and \citet{Lopin
etal.14}  can not be caused by the radial component of the
equilibrium magnetic field~$B_r(r,z)$ and the main reason is
related to equation of motion (\ref{spruit motion}), which seems
to be not quite reasonable (see Section~3).

\section{Discussion and conclusions}

In the present work we check the statement of \citet{Lopin
etal.14} that Sprit's approach based on the phenomenological
equation of motion can be used for description of kink waves of
thin magnetic flux tubes if we take into account the radial
component of the tube in the induction equation. As it follows
from the our results this statement is not quite correct since the
main problem is related to the phenomenological equation of motion
which  can not be used for the description of the external force
outside the tube. This conclusion also suggests that the the
phenomenological equation of motion can not be used to describe
the dynamics of thin flux tubes.

Following \citet{Lopin&Nagorny13} we have confirmed that the low
frequency kink waves of thin magnetic flux tubes are cutoff free
modes. In particularly, this suggests that these modes generated
in the photosphere by convective motions can be responsible for
heating and accelerating the fast wind in spite of quite strong
reflection in the transition region~\citep{Tsap06, Soler etal.17}.
Also they can be very useful for plasma diagnostics and can
support the decayless transverse oscillations of coronal
loops~\citep{Nistico etal.13}. However,  the thin flux tube
approximation has serious restrictions therefore the investigation
of the kink mode propagation in the tubes with more complex
magnetic configuration is required.


\begin{thebibliography}{}

\vspace{-3mm}
\bibitem[\protect\citeauthoryear{{Cheng}}{1992}]{Cheng92}
--- {\it Cheng J.}
Equations for the motion of an isolated thin magnetic flux tube
// Astron. Astrophys. V.~264. N~1. P.~243--248. 1992.

\vspace{-3mm}
\bibitem[\protect\citeauthoryear{{Choudhuri}}{1990}]{Choudhuri90}
--- {\it Choudhuri A.R.}
A correction to Spruit's equation for the dynamics of thin flux tubes
// Astron. Astrophys. V.~239. N~1--2. P.~335--339. 1990.

\vspace{-3mm}
\bibitem[\protect\citeauthoryear{{De Pontieu \etal}}{2001}]{De Pontieu et al.01}
--- {\it De Pontieu B., Martens P.C.H., Hudson H.S.}
Chromospheric Damping of Alfv\'{e}n Waves
// Astrophys. J. V.~558. N~2. P.~859--871. 2001.

\vspace{-3mm}
\bibitem[\protect\citeauthoryear{{Fan}}{2009}]{Fan.09}
--- {\it Fan Y.}
Magnetic Fields in the Solar Convection Zone
// Liv. Rev. Solar Phys. V.~6. N~1. id.4. 96~P. 2009.

\vspace{-3mm}
\bibitem[\protect\citeauthoryear{{Fan, Fisher, and McClymont}}{1994}]{Fan etal.94}
--- {\it Fan Y., Fisher G.H., McClymont A.N.}
Dynamics of emerging active region flux loops
//  Astrophys. J. V.~436. N~2. P.~907--928. 1994.

\vspace{-3mm}
\bibitem[\protect\citeauthoryear{{Ferriz-Mas} and {Sch\"ussler}}{1989}]{Ferriz-Mas at al.89}
--- {\it Ferriz-Mas A., Sch\"ussler M.A.V.} //
Dynamics of magnetic flux concentrations - The second-order thin flux tube approximation
Astron. Astrophys. V.~210. N~1--2. P.~425--432. 1989.

\vspace{-3mm}
\bibitem[\protect\citeauthoryear{{Fujimura} and {Tsuneta}}{2009}]{Fujimura&Tsuneta09}
--- {\it Fujimura D., Tsuneta S.}
Properties of Magnetohydrodynamic Waves in the Solar Photosphere Obtained with Hinode
// Astrophys. J. V.~702. N~2. P.~1443--1457. 2009.

\vspace{-3mm}
\bibitem[\protect\citeauthoryear{{Gelfreikh \etal}}{2004}]{Gelfreikh etal.04}
--- {\it Gelfreikh G.B., Tsap Yu.T., Kopylova Yu.G. et al.}
%, Goldvarg, T.B., Nagovitsyn, Yu.A., Tsvetkov, L.I.:
Variations of Microwave Emission from Solar Active Regions
// Astron. Letters V.~30. 489--495. 2004.

\vspace{-3mm}
\bibitem[\protect\citeauthoryear{{Hollweg}}{1981}]{Hollweg81}
--- {\it Hollweg J.V.}
Alfv\'{e}n waves in the solar atmosphere. II - Open and closed magnetic flux tubes
// Solar Phys. V.~70. P.~25--66. 1981.

\vspace{-3mm}
\bibitem[\protect\citeauthoryear{{Hollweg}}{1984}]{Hollweg84}
--- {\it Hollweg J.V.}
Resonances of coronal loops
// Astrophys. J. V.~277. P.~392--403. 1984.

\vspace{-3mm}
\bibitem[\protect\citeauthoryear{{Howard} and {Stenflo}}{1972}]{Howard&Stenflo72}
--- {\it Howard R., Stenflo J.O.}
On the Filamentary Nature of Solar Magnetic Fields
// Solar Phys. V.~22. N~2. P.~402--417. 1972.

\vspace{-3mm}
\bibitem[\protect\citeauthoryear{{Jess \etal}}{2017}]{Jess et al.17}
--- {\it Jess D.B.,  Van Doorsselaere T., Verth G. et al.}
An Inside Look at Sunspot Oscillations with Higher Azimuthal Wavenumbers
// Astrophys. J. V.~842. N~1. id.~59. 9~p. 2017.

\vspace{-3mm}
\bibitem[\protect\citeauthoryear{{Ji \etal}}{2012}]{Ji et al.12}
--- {\it Ji H., Cao W., Goode P.R.}
Observation of Ultrafine Channels of Solar Corona Heating
// Astrophys. J. Letters. V.~750. N~1. id.~L25. 5~p. 2012

\vspace{-3mm}
\bibitem[\protect\citeauthoryear{{Longcope and Klapper}}{1997}]{Longcope&Klapper97}
--- {\it Longcope D.W., Klapper I.}
Dynamics of a Thin Twisted Flux Tube
// Astrophys. J. V.~488. N~1. P.~443--453. 1997.

\vspace{-3mm}
\bibitem[\protect\citeauthoryear{{Leake \etal}}{2005}]{Leake etal.05}
--- {\it Leake J.E., Arber T.D., Khodachenko M.L.}
Collisional dissipation of Alfv\'{e}n waves in a partially ionised solar chromosphere
// Astron. Astrophys. V.~442. N~3. P.~1091--1098. 2005.

\vspace{-3mm}
\bibitem[\protect\citeauthoryear{{Lopin} and {Nagorny}}{2013}]{Lopin&Nagorny13}
--- {\it Lopin I., Nagorny I.}
Conditions for Transverse Waves Propagation along Thin Magnetic Flux Tubes on the Sun
// Astrophys. J. V.~774. N~2. id.~121. 5~p. 2013.

\vspace{-3mm}
\bibitem[\protect\citeauthoryear{{Lopin \etal}}{2014}]{Lopin etal.14}
--- {\it Lopin I.P., Nagorny I.G., Nippolainen E.}
Kink Wave Propagation in Thin Isothermal Magnetic Flux Tubes
// Solar Phys. V.~289. N~8. P.~3033--3041. 2014.

\vspace{-3mm}
\bibitem[\protect\citeauthoryear{{Lopin} and {Nagorny}}{2017}]{Lopin&Nagorny17}
--- {\it Lopin I., Nagorny I.}
Kink Waves in Thin Stratified Magnetically Twisted Flux Tubes
// Astrophys. J. V.~840. N~1. id.~26. 7~p. 2017.

\vspace{-3mm}
\bibitem[\protect\citeauthoryear{{Moreno-Insertis, Ferriz-Mas, and Schl\"ussler}}{1996}]{Moreno-Insertis etal.96}
--- {\it Moreno-Insertis F., Ferriz-Mas A., Schl\"ussler M.}
Enhanced inertia of thin magnetic flux tubes
// Astron. Astrophys. V.~312. P.~317--326. 1996.

\vspace{-3mm}
\bibitem[\protect\citeauthoryear{{Morton \etal}}{2015}]{Morton et al.15}
--- {\it Morton R.J., Tomczyk S., Pinto R.}
Investigating Alfv\'{e}nic wave propagation in coronal open-field regions
// Nat. Comm. V.~6. id.~7813. 2015.

\vspace{-3mm}
\bibitem[\protect\citeauthoryear{{Musielak} and {Ulmschneider}}{2001}]{Musielak&Ulmschneider01}
--- {\it Musielak Z.E., Ulmschneider P.}
Excitation of transverse magnetic tube waves in stellar convection zones. I. Analytical approach
// Astron. Astrophys. V.~370. P.~541--554. 2001.

\vspace{-3mm}
\bibitem[\protect\citeauthoryear{{Nistic\'o \etal }}{2013}]{Nistico etal.13}
--- {\it Nistic\'o G., Nakariakov V.M., Verwichte E.}
Decaying and decayless transverse oscillations of a coronal loop
// Astron. Astrophys. V.~552. id.~A57. 6~p. 2013.

\vspace{-3mm}
\bibitem[\protect\citeauthoryear{{Osin, Volin, and Ulmschneider}}{1999}]{Osin etal.99}
--- {\it Osin A., Volin S., Ulmschneider P.}
Propagation of nonlinear longitudinal-transverse waves along magnetic flux tubes in the solar atmosphere. III. Modified equation of motion
// Astron. Astrophys. V.~351. P.~359--367. 1999.

\vspace{-3mm}
\bibitem[\protect\citeauthoryear{{Roberts} and {Webb}}{1978}]{Roberts&Webb78}
--- {\it Roberts B., Webb A.R.}
Vertical motions in an intense magnetic flux tube
// Solar Phys. V.~56. P.~5--35. 1978.

\vspace{-3mm}
\bibitem[\protect\citeauthoryear{{Ryutov and Ryutova}}{1976}]{Ryutov&Ryutova76}
--- {\it Ryutov D.D.,  Ryutova M.P.}
Sound oscillations in a plasma with ``magnetic filaments''
// Sov. Phys. JETP. V.~43. N~3. P.~491--497. 1976

\vspace{-3mm}
\bibitem[\protect\citeauthoryear{{Routh \etal}}{2007}]{Routh etal.07}
--- {\it Routh S., Musielak Z.E., Hammer R.}
Conditions for Propagation of Torsional Waves in Solar Magnetic Flux Tubes
// Solar Phys. V.~246. N~1. P.~133--143. 2007.

\vspace{-3mm}
\bibitem[\protect\citeauthoryear{{R\"uedi \etal}}{1992}]{Ruedi etal.92}
--- {\it R\"uedi  I., Solanki S.K., Livingston W., Stenflo J.O.}
Infrared lines as probes of solar magnetic features. III - Strong and weak magnetic fields in plages
// Astron. Astrophys. V.~263. N~1--2. P.~323--338. 1992.

\vspace{-3mm}
\bibitem[\protect\citeauthoryear{{Sharykin} and {Kosovichev}}{2014}]{Sharykin&Kosovichev14}
--- {\it Sharykin I.N., Kosovichev A.G.}
Fine Structure of Flare Ribbons and Evolution of Electric Currents
// Astrophys. J. Letters. V.~788. N~1. id.~L18. 7~p. 2014.

\vspace{-3mm}
\bibitem[\protect\citeauthoryear{{Soler \etal}}{2017}]{Soler etal.17}
--- {\it Soler R., Terradas J., Oliver R., Ballester J.L.}
Propagation of Torsional Alfv\'{e}n Waves from the Photosphere to the Corona: Reflection, Transmission, and Heating in Expanding Flux Tubes
Astrophys. J. V.~840. id.20. 18~p. 2017.

\vspace{-3mm}
\bibitem[\protect\citeauthoryear{{Spruit}}{1981}]{Spruit81}
--- {\it Spruit H.C.}
Motion of magnetic flux tubes in the solar convection zone and chromosphere
// Astron. Astrophys. V.~98. P.~155--160. 1981.

\vspace{-3mm}
\bibitem[\protect\citeauthoryear{{Stenflo}}{2011}]{Stenflo11}
--- {\it Stenflo J.O.}
Collapsed, uncollapsed, and hidden magnetic flux on the quiet Sun
// Astron. Astrophys. V.~529. A42. 20~p. 2011.

\vspace{-3mm}
\bibitem[\protect\citeauthoryear{{Tsap} and {Kopylova}}{2001}]{Tsap et al.01}
--- {\it Tsap Yu.T., Kopylova Yu.G.}
Acoustic Damping of Fast Kink Oscillations of Coronal Loops
// Astron. Letters. V.~27. N~11. P.~737--744. 2001.

\vspace{-3mm}
\bibitem[\protect\citeauthoryear{{Tsap}}{2006}]{Tsap06}
--- {\it Tsap Y.T.}
On the penetration of Alfv\'{e}n waves from the chromosphere into the corona
// Proc. IAU Symp. N~233, Solar Activity and its Magnetic Origin.  Eds. Bothmer V. and Hady A.A.
Cambridge Univ. Press, Cambrige, p.253--254. 2006.

\vspace{-3mm}
\bibitem[\protect\citeauthoryear{{Weber and Browning}}{2016}]{Weber&Browning16}
--- {\it Weber M.A., Browning M.K.}
Modeling the Rise of Fibril Magnetic Fields in Fully Convective Stars
//  Astrophys. J. V.~827. id.95. 20~p. 2016.

\end{thebibliography}
\end{document}